\documentclass{ifacconf}

\usepackage{graphicx}      
\usepackage{natbib}        
\usepackage{amsmath} 
\usepackage{amssymb}  

\usepackage{pgfplots}
\pgfplotsset{compat=newest}
\usepackage{placeins}
\usepackage{dblfloatfix}

\usepackage[hyphens]{url}

\usepackage{graphicx}      
\usepackage{tikz}          
\usepackage{tikzscale}
\usepackage{pgfplots}      
\usepackage{subcaption}    
\usepackage{caption} 

\usepackage{array}
\usepackage{booktabs}
\usepackage{multirow}
\usepackage{csquotes}
\usepgfplotslibrary{statistics}

\makeatletter
\let\old@ssect\@ssect 
\makeatother

\usepackage{hyperref}
\hypersetup{
	colorlinks=false,
	linkbordercolor = {white},
	pdfborder={0 0 0},
}
\makeatletter
\def\@ssect#1#2#3#4#5#6{%
	\NR@gettitle{#6}
	\old@ssect{#1}{#2}{#3}{#4}{#5}{#6}
}
\makeatother

\usepackage{fancyheadings}

\usepackage{multicol}
\usepackage{fancyhdr}
\fancypagestyle{firstpage}{
	\fancyhf{}
	
	\fancyhead[R]{\vspace{-0.5mm}  \copyright~2025 Balint Varga, Lars Fischer, Levente Kovacs\\
		\vspace{0.5mm}
		This work has been accepted to IFAC for publication \\
		under a Creative Commons
		Licence CC-BY-NC-ND \\ 
		\vspace{0.5mm}} 
}

\begin{document}
\begin{frontmatter}


\title{Personalized and Demand-Based Education Concept: Practical Tools for Control Engineers\thanksref{finsup}} 
\thanks[finsup]{This publication was written within the framework of the KAMO: Karlsruhe Mobility High Performance Center (\href{www.kamo.one}{www.kamo.one}).}

\author[First]{Balint Varga},
\author[Second]{Lars Fischer} 
\author[Third]{Levente Kovacs}

\address[First]{Institute of Control Systems, Karlsruhe Institute of Technology, 76131 Karlsruhe, Germany, e-mail: balint.varga2@kit.edu.}
\address[Second]{FZI, Research Center for Information Technology, 76131 Karlsruhe, Germany}
\address[Third]{Department of Biomatics and Artificial Intelligence, John von Neumann Faculty of Informatics, and Physiological Controls Research Group, University Research and Innovation System, Obuda University, 1034, Budapest, Hungary}

\begin{abstract}                
This paper presents a personalized lecture concept using \textit{educational blocks }and its demonstrative application in a new university lecture. Higher education faces daily challenges: deep and specialized knowledge is available from everywhere and accessible to almost everyone. University lecturers of specialized master courses confront the problem that their lectures are either too boring or too complex for the attending students. Additionally, curricula are changing more rapidly than they have in the past 10--30 years.
The German education system comprises different educational forms, with universities providing less practical content. Consequently, many university students do not obtain the practical skills they should ideally gain through university lectures.
Therefore, in this work, a new lecture concept is proposed based on the extension of the just-in-time teaching paradigm: Personalized and Demand-Based Education. This concept includes: 1) an initial assessment of students' backgrounds, 2) selecting the appropriate educational blocks, and 3) collecting ongoing feedback during the semester. The feedback was gathered via Pingo, ensuring anonymity for the students. Our concept was exemplarily tested in the new lecture "Practical Tools for Control Engineers" at the Karlsruhe Institute of Technology. The initial results indicate that our proposed concept could be beneficial in addressing the current challenges in higher education.
\end{abstract}

\thispagestyle{firstpage}


\end{frontmatter}
\thispagestyle{firstpage}

\section{Introduction}
Higher education plays an important role in shaping the skills and competencies of future professionals, driving innovation, and advancing our society. 
However, it is confronted by many challenges that can undermine its impact and relevance: see \cite{2011_bounded_rationality} or \cite{2022_rethink_education}.
The vast availability of specialized knowledge is among these challenges, which (while democratizing information access) leads to information overload and heightened competition among institutions to provide distinctive and impactful education. In the age of generative artificial intelligence, lecturers and university professors are confronted with the task of not only delivering content but also ensuring that their teaching remains relevant and engaging in an environment where information is readily accessible to everyone.

In addition to technological advancements, there is a growing concern regarding the attention spans of young students \cite{bunce2010long,bradbury2016attention}. Maintaining student engagement is increasingly difficult, as distractions are plentiful, and students often prefer interactive and multimedia content over traditional lecture formats \cite{schwerdt2011traditional}. Consequently, university lectures must add significant value compared to online tutorials or prerecorded lecture videos to justify students' time and effort in attending them. Furthermore, monotonous lectures can easily lose students' attention, while overly difficult content lowers motivation. Academic programs now evolve rapidly to keep pace with technology, industry demands, and emerging fields. Consequently, flexible, innovative teaching methods are crucial for meeting changing educational needs and student expectations, see \cite{Rossiter2023}. 

Within the German education system, the diversity of educational formats\footnote{These include \textit{Universit\"aten, Hochschulen und Duale Hochschulen}, which can be translated as research universities, Universities of Applied Sciences, and Dual Universities of Cooperative Education. For more information, the reader is referred to \cite{hochschulkompass2023}} offers a range of pedagogical approaches. Universities often prioritize theoretical teaching content at the expense of practical training. This prioritization can lead to master's and doctoral students frequently having difficulties transitioning from the academic world to working in industry. This is due to university curricula often focusing on theoretical problems, which results in graduates who -- despite possessing deep theoretical understanding -- lack the hands-on experience necessary to apply their knowledge effectively to real-world problems.

\pagestyle{empty}
Addressing these challenges requires a paradigm shift in lecture design and delivery, which is also the goal of the Technical Committees of the IFAC and IEEE societies; see \cite{2019_TechnicalCommitteeControl_rossiter,Rossiter2023,2023_TC_IFAC}. Therefore, this paper proposes a personalized lecture concept that uses \textit{educational blocks} to tailor content to individual student needs. Based on the Just-in-Time Teaching (JiTT) paradigm, our proposed {Personalized and Demand-Based Education} (PDBE) concept aims to enhance lecture engagement and efficacy by customizing content delivery based on graduate students' backgrounds and ongoing feedback, focusing on practical topics relevant for industry applications. 

Our PDBE concept comprises three key components: (1) an initial evaluation of students' backgrounds to understand their prior knowledge, (2) the selection of appropriate educational blocks that align with both curricular objectives and individual student profiles, and (3) the continuous collection of feedback throughout the semester to adjust and refine the teaching approach dynamically. To facilitate anonymous feedback, we use the Pingo platform hosted by Karlsruhe Institute of Technology (KIT), which ensures student anonymity and encourages candid responses. We implemented our novel PDBE concept in the \textit{Practical Tools for Control Engineers}\footnote{The lecture website is available at \url{https://www.irs.kit.edu/english/Lectures_4827.php}.} course at the KIT. This application served as an exemplary case to test the efficacy of our framework in a real-world educational setting. 

The remainder of this paper is structured as follows: Section \ref{sec:theoretical_background} reviews the relevant literature on JiTT paradigms and personalized education concepts. Section \ref{sec:Proposed_Concept} details the methodology of our study, including the implementation of the proposed concept. Section \ref{sec:results} presents the results of our preliminary findings, while Section \ref{sec:Discussion} discusses the implications and limitations of these results. Finally, Section \ref{sec:summary} concludes the paper and outlines directions for future research.

\vspace*{-1.5mm}
\section{Theoretical Background}
\label{sec:theoretical_background}
Due to recent challenges, many engineering educators struggle to keep pace with industry demands and interdisciplinary competencies (\cite{2011_bounded_rationality,2022_rethink_education}).
Therefore, recent research focuses on improving teaching methodologies in higher education, particularly in engineering and technology domains. 
This section gathers the most applicable work for this paper into two groups, providing a concise literature overview.
\vspace*{-1mm}
\subsection{Just-in-Time Teaching Concepts}
\vspace*{-1mm}
One such innovative approach is the JiTT, a pedagogical strategy that synchronizes pre-class preparation with in-class activities (\cite{ChengXu2020,FoxDoherty2021}). 
Students complete brief assignments online before coming to class, allowing instructors to adapt the upcoming session based on the submitted work. This method fosters active learning by addressing misconceptions and knowledge gaps immediately (\cite{2015_UsingFlippedClassroom_jonssonc,2017_JustintimeTeachingImproves_perez-poch,2024_EffectsJustintimeInquiry_fukuda}). 
It also encourages students to stay engaged, as they see that their contributions shape the class activities. By shifting part of the content delivery and assessment outside the classroom, JiTT opens up valuable face-to-face time for deeper exploration (\cite{TuckerGriffin2017,GuptaLee2021}). Effective JiTT implementation can improve student motivation and performance, as demonstrated in computer science fundamentals (\cite{2017_JustintimeTeachingImproves_perez-poch}). 
The literature suggests that JiTT helps maintain students' attention and motivation, especially when used alongside interactive classroom strategies like group work or live simulations.
\vspace*{-1mm}
\subsection{Personalized, Practical and Competence-based Teaching Concepts}
\vspace*{-1mm}
In parallel, personalized or demand-based education concepts highlight that students enter courses with varying backgrounds, skills, and learning styles \cite{SmithEtAl2019,PeterDeVries2018}. The theoretical foundation of personalized education suggests instructional blocks and content sequencing should adapt to each student's prior knowledge, emphasizing differentiated instruction and scaffolding where necessary. Research indicates that combining personalized lecture content with regular students' feedback better aligns curricular goals, enhancing comprehension and retention of core concepts \cite{ClarosDuart2021,ChangEtAl2022}. Personalized and Practical approaches are particularly valuable in fast-evolving fields such as control engineering, where technologies and methodologies develop rapidly (\cite{OrrEtAl2022}).

A consistent finding across studies emphasizes the critical role of practical, hands-on experiences for developing problem-solving skills alongside theoretical understanding (\cite{AlonsoEtAl2019,BogunovicTheis2022}). Project-based learning, simulation labs, and real-world case studies are frequently cited as effective methods to bridge theory-practice gaps. These approaches gain additional potential when they are coupled with adaptive feedback systems that helps instructors to analyze student performance in virtual labs or mini-projects \cite{UmezawaEtAl2018,MakranskyPetersen2019}, subsequently adjusting content delivery, scheduling correction, or providing alternative resources. For broader perspectives, see comprehensive reviews in \cite{2018_SurveyGoodPractice_rossiter,Rossiter2023}.

The ``community roadmap'' \cite{Rossiter2023} outlines how evolving societal priorities---including sustainability, infrastructure development, and digital transformation---necessitate broadening research agendas, training programs, and flexible course architectures. It further advocates for novel learning formats (e.g., modular content, micro-courses) to address complex societal challenges.

Building on this literature review, this paper proposes an adapted implementation of the JiTT framework for graduate control engineering education at a German university, designed to prepare students for real-world engineering problem-solving including software engineering components. The proposed concept uses education blocks similar to the learner-centric design principles \cite{ClarosDuart2021}.

\vspace*{-0.5mm}
\section{The Personalized Demand-Based Education Concept}
\label{sec:Proposed_Concept}
\begin{figure}[t]
	\begin{center}
		\includegraphics[width=0.82\linewidth]{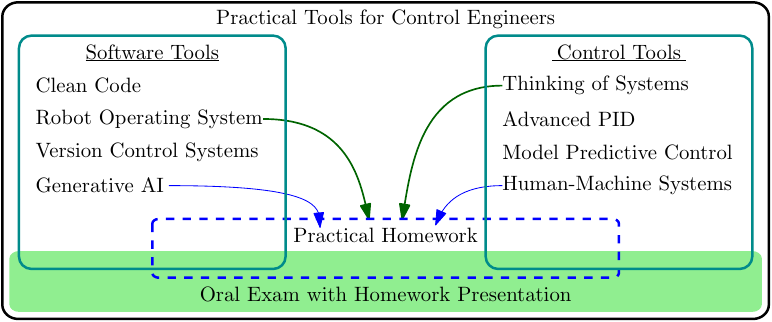}  
		\caption{The educational blocks of the lecture \textit{Practical Tools for Control Engineers} as the current implementation of our PDBE concept} 
		\label{fig:blocks_for_demand}
	\end{center}
\end{figure}

\vspace*{-0.5mm}
This section introduces the personalized sets of education blocks for our proposed PDBE concept, which are divided into two main categories: \textit{Software Tools} and \textit{Control Tools}. The exemplary components for the first implementation during the semester are given in Fig.~\ref{fig:blocks_for_demand}. The students' feedbacks guide the selection of these educational blocks, ensuring that most of the students receive content tailored to their background. The usage of these two main blocks are justified by the literature indicating the benefit of the competence-based teaching concepts. 
Thus, through this concept, control engineering students can gain a foundational set of tools to effectively solve industry-relevant control and systems engineering problems.

\vspace*{-0.95mm}
\subsection{Demand Identification and Implementation}
\vspace*{-0.95mm}
Central to our approach is a demand identification using Pingo Questions at the lectures, which is feedback procedure that tailors lecture content.
\begin{itemize}
	\item[1] \textbf{Preliminary Student Assessment:} At the beginning of the course, students participate in a needs assessment through a questionnaire focused on their prior experience with software tools and control theory. This initial assessment affects our instructional approach, allowing us to provide targeted support and ensure that the course content is both accessible and engaging for all students.
	
	\item[2] \textbf{Identification of Knowledge Gaps:} Using the questionnaire results, the instructor has to determine which software or control topics are most relevant for students. Some of them may need extra time on coding basics, while others might benefit from more advanced control material.
	
	\item[3] \textbf{Selection of Education Blocks:} A customized selection of software and control blocks happens based on the identified students knowledge gap. This ensures lectures remain engaging for most students despite of the varying backgrounds.
	
	\item[4] \textbf{Continuous Semester Feedback:} After each block, students submit anonymous feedback on the lecture. The instructor uses this information to clarify confusing points or add examples in subsequent sessions. Future content is adjusted to address their needs, making the learning process flexible and responsive.
\end{itemize}

In the upcoming subsection, the necessary tools are presented.
\vspace*{-0.95mm}
\subsection{First Educational Block: Software Tools}
\vspace*{-1.25mm}
Students first explore essential software and programming topics central to modern control engineering. The focus lies on writing clean, maintainable code, using version control to manage collaborative work, and understanding how to apply different programming languages or development environments. For instance, Python is emphasized for its readability and extensive library support, while C/C++ or MATLAB may be chosen for performance-critical tasks or advanced mathematical modeling.

\vspace*{-0.95mm}

Students additionally learn simulation and design techniques that enable them to verify control strategies and optimize system parameters before real-world deployment. Through practical mini-examples -- such as simulation of robotic applications and simple industrial processes -- students discover how these tools integrate with real-world control engineering tasks. The goal is to establish reliable software foundations, empowering learners to confidently select and apply appropriate technologies in diverse scenarios.

\vspace*{-1.2mm}
\subsection{Second Educational Block: Control Tools}
\vspace*{-0.75mm}
Once students acquire the necessary software skills, they continue with practical control topics covering advanced Proportional Integral Derivative (PID) controller and Model Predictive Control (MPC). 

\vspace*{-0.75mm}

Since PID controller is still relevant for most of the industrial applications, its extensions were presented (Anti Wind-Up, Smith Predictor, MIMO PID). Model Predictive Control (MPC) is introduced in simplified linear form, demonstrating how stability concepts and constraint handling operate through simulation examples. Students also examine human-robot interaction systems where control strategies must ensure safety and efficiency.

\vspace*{-1mm}
\subsection{Implementation of the Proposed Concepts in the Practical Tools for Control Engineers Lecture}
\vspace*{-1mm}
The software and control tools discussed in the previous subsections belong to competence-based teaching methods. By using student feedback for personalization, we managed to increase student engagement during lectures. Furthermore, we used the interactive Pingo tool to provide an additional value to our work, which are presented in the following subsections.
\vspace*{-1mm}
\subsubsection{Engagement and Motivation Tools} 
\phantom{a}\newline
The Pingo system was utilized to conduct short, in-class quizzes throughout the course. These quizzes provided students with opportunities to test their understanding of newly learned concepts. Similar to the JiTT approach, by providing immediate feedback on knowledge gaps, we could address misconceptions and areas of confusion promptly. This real-time feedback enabled identification of topics requiring further clarification, allowing subsequent lecture adjustments. To increase engagement of these activities, extra credit for the exam could be collected. To ensure anonymity, all in the Pingo test participating students received these bonus points if 66\% of them answered 66\% of the questions correctly. 	
\vspace*{-1mm}
\subsubsection{Additional Implementation Considerations}  
\phantom{a}\newline
In addition to quick quizzes, we included presentations of practical research projects, see e.g. \cite{varga2023intention}. These sessions showcased our institute's research initiatives, emphasizing the practical relevance of lecture content. Specifically, projects focused on developing advanced human-machine interaction systems demonstrated real-world applications of theoretical concepts, see \cite{2024_sean_roman}. 
\vspace*{-1mm}
\subsubsection{Considerations for the Homework}  
\phantom{a}\newline
Students were required to complete a comprehensive homework assignment as a prerequisite for the oral exam. To simulate real-world experiences, they faced challenges typical of open-source projects, including software installation difficulties\footnote{While tools like ROS2, Python, and CasADi are well-documented, students encountered common issues such as dependency conflicts and platform-specific configurations.}. For this task, the students developed model predictive controllers suitable for modeling human-machine interactions using the framework presented in~\cite{varga2023cooperative}.

\vspace*{-2.2mm}
\section{Results}
\label{sec:results}
\vspace*{-2.2mm}

\begin{table}[t!]
	\centering
	\begin{tabular}{|l|c|}
		\hline
		\textbf{Range (Lines of Code)} & \textbf{Occurrences} \\ \hline
		0--100 & 4 \\ \hline
		101--1000 & 4 \\ \hline
		1001+ & 5 \\ \hline
	\end{tabular}
	\vspace*{1mm}
	\captionsetup{justification=centering, width=\linewidth} 
	\caption{Answers for the question \textit{How long was}\\ 
		\textit{your longest project?}} 
	\label{fig:background_students_line_of_code}
\end{table}
For the first time, the course “Practical Tools for Control Engineers” was offered in the winter term 2024/25 at KIT. It was attended regularly by 18–22 students throughout the semester, which consisted of 15 weeks of 1.5-hour lectures. The course was organized into three parts: a 6-week software tools block, a 7-week control tools block, and a 2-week block dedicated to preparing for homework and the exam.
The following presents the questions raised during the first lecture along with the students’ answers.

\begin{figure}[t!]
	\centering
	\begin{tikzpicture}
		\begin{axis}[
			ybar,
			symbolic x coords={0, 1, 2, 3, 4, 5, 9, 11, 15, 30, 60},
			xtick=data,
			ylabel={Occurrences},
			ymin=0, ymax=4.2,
			bar width=15pt,
			height=4.15cm,      
			width=0.99\linewidth,  
			grid=both,
			]
			\addplot coordinates {(0,1) (1,2) (2,2) (3,4) (4,1) (5,2) (9,1) (11,1) (15,1) (30,1) (60,2)};
		\end{axis}
	\end{tikzpicture}
	\caption{Background Information: Answers for \textit{With how many peers have you worked together (number of project members, open-source vs. private projects)?}}
	\label{fig:background_students_peers}
\end{figure}
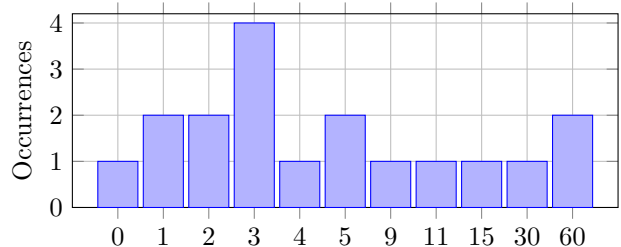

\begin{itemize}
	\item[Q1] How long was your longest project? (lines of code)		
	\item[A1] Table.~\ref{fig:background_students_line_of_code}
	\item[Q2] Which Programming Languages do you know?
	\item[A2] 
	\begin{itemize}
		\item Python (12)
		\item Matlab (9)
		\item C (7)
		\item C++ (5)
		\item C\# (2)
	\end{itemize}
	\item[Q3] In what kind of applications have you used your programming knowledge?
	\item[A3] Table~\ref{fig:stunendt_applications}
\item[Q4] With how many peers have you worked together (number of project members, open-source vs. private projects)?
\item[A4] Fig.~\ref{fig:background_students_peers}
\end{itemize}

\begin{table}[ht]
	\centering
	\begin{tabular}{l c}
		\hline
		\textbf{Group} & \textbf{Count} \\
		\hline
		Web-development & 4 \\
		Microcontroller & 7 \\
		Control tasks & 10 \\
		Simulation & 2 \\
		Autonomous driving & 1 \\
		AI-Tooling/Database & 10 \\
		Robotics & 2 \\
		\hline
	\end{tabular}
	\vspace*{1mm}
	\captionsetup{justification=centering, width=\linewidth} 
	\caption{Student Background on applications}
	\label{fig:stunendt_applications}
\end{table}

	%
	

\begin{figure*}[t]
	\centering
	\begin{subfigure}[t]{0.48\textwidth}
		\centering
		\includegraphics[width=0.75\linewidth]{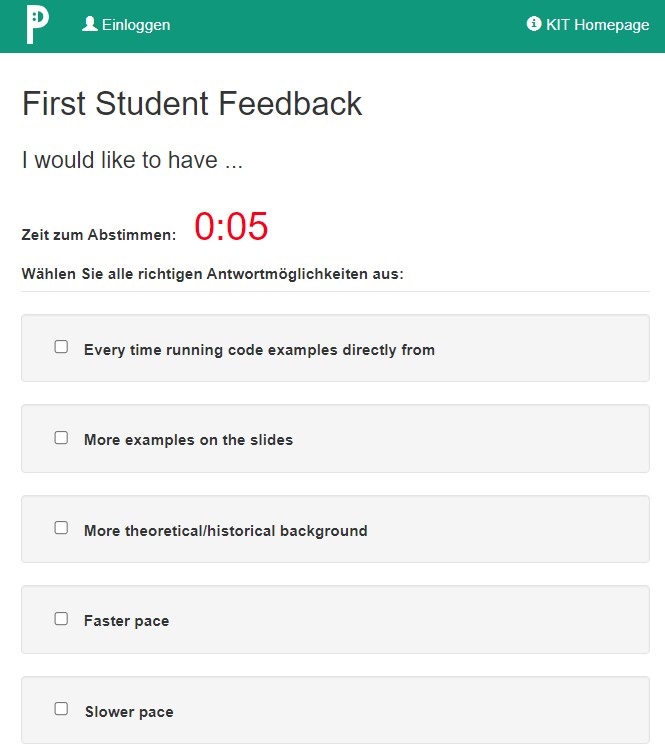}
		\caption{Student Feedback Question Round Using Pingo Hosted by KIT}
		\label{fig:feedback_students_pingo_view}
	\end{subfigure}
	\hfill
	\begin{subfigure}[t]{0.45\textwidth}
		\centering
		\begin{tikzpicture}
			\begin{axis}[
				ybar,
				symbolic x coords={A, B, C, D, E},
				xtick=data,
				ylabel={Number of Answers},
				xlabel={Answer Option},
				ymin=0, ymax=15,
				bar width=25pt,  
				height=6.5cm,      
				width=0.99\linewidth,  
				grid=both,
				tick label style={font=\footnotesize},
				label style={font=\small},
				legend style={
					at={(0.5,-0.35)},  
					anchor=north,
					legend columns=1,
					legend cell align=left,
					draw=none,
					font=\scriptsize
				}
				]
				\addplot coordinates {(A,14) (B,6) (C,3) (D,2) (E,6)};
			\end{axis}
		\end{tikzpicture}
		\caption{Feedback results: \emph{"I would like to have..."} \\
			A: Code execution every time \; B: More slide examples \\
			C: More theory/history \; D: Faster pace \;	E: Slower pace}
		\label{fig:feedback_bar_chart}
	\end{subfigure}
	\caption{Pingo view of the student questionnaire (left side) and the results (right side) indicating that the presentation of the code examples in the previous lectures was not sufficient}
	\label{fig:feedback_overview}
\end{figure*}

Students' background identification showed that they have some relevant experience working in a team and handling larger code bases. On the other hand, some of them have never worked in a team and have only a little coding experience. This heterogeneous background had to be taken into account during the fine-tuning of the lecture content.

Fig.~\ref{fig:feedback_overview} shows the students' Pingo results and the outcome of a feedback question. It turned out that they preferred to carry out the example codes together during the lecture instead of doing them as after-class assignments. This feedback was taken into account during the semester, and dedicated time slots were planned for carrying out the lecture examples and demonstrating the working principles of the code examples.

During the semester, seven in-class mini quizzes were conducted. Only the students who engaged in lecture-to-lecture learning during the semester could answer them. Only once, 66\% of the questions were answered correctly by 66\% of the participating students, indicating that most students did not prepare for the lectures.

Evaluation Office of the Quality Management Department at KIT offers a central teaching evaluation, which provides a standardized feedback system for lecturers. 
"The teaching evaluation at KIT enables students to anonymously provide feedback on courses. Each evaluation questionnaire includes six mandatory questions, which are used to calculate the Teaching Quality Index for a course. The questions include
\begin{itemize}
	\item the evaluation of the content of the course,
	\item the evaluation of the teaching quality of the full-time and part-time lecturers,
	\item the assessment of the organization and supervision of the course
	\item the assessment of student commitment in connection with the course,
	\item the assessment of the infrastructure,
	\item the overall assessment of the course.
\end{itemize}
The lecture achieved above-average results, which indicate the relevance and the suitability of the applied PDBE concept.

\vspace*{-2.2mm}
\section{Discussion}
\label{sec:Discussion}
\vspace*{-2.2mm}
The main limitation of our concept was the limited resources during the preparation of the lecture. Since the education blocks were not prepared for all possible scenarios, they had to be extended during the semester to respond in accordance with the students' feedback. Furthermore, there is no baseline for our results. Comparing different lectures does not provide a sufficient benchmark for rigorous evaluation. Consequently, additional structured studies are needed to identify best practices and consolidate consistent data.

We also did not fully address the question of scaling the concept to larger classes. Our current findings are confined to smaller groups, and significant adjustments in resources and infrastructure would be required to confirm its feasibility and effectiveness on a larger scale.

We are planning to address these limitations. Furthermore, the question of how the practical implementation of more advanced mathematical concepts, like \cite{2021_OrdinalPotentialDifferential_varga,varga2024toward}, 
could be integrated into a seminar-like extension of our current lecture will be answered.
\vspace*{-2.25mm} 
\section{Conclusion} \label{sec:summary}
\vspace*{-2.25mm}
This paper presents the application of a novel educational concept, Personalized and Demand-Based Practical Education, utilizing educational blocks to enhance lecture engagement and effectiveness. Furthermore, we apply the concept to the new lecture Practical Tools for Control Engineers at the KIT. The preliminary results showed that the concept is applicable to a lecture at a university and is accepted by the students. 

\vspace*{-1.25mm}

In our future work, a) we aim to adapt the concept for seminar-based courses, which can emphasize more interactive and discussion-oriented learning over traditional lectures. Furthermore, we are planning to include additional transportation research topics, like e.g. platooning \cite{2018_Platooning_HasanzadeZonuzy} or automotive sensor fusion algorithms \cite{2022_sesorFusion_Linden}. Seminars and additional research topics can better prepare students for practical laboratory work at our institute by developing essential skills like problem-solving and collaboration.

\section*{Acknowledgments}
\vspace*{-1mm}
This publication was written within the framework of the KAMO: Karlsruhe Mobility High Performance Center (\url{www.kamo.one}), the association of Karlsruhe’s institutions for research, education, and transfer in the field of innovative mobility and logistics solutions.
\vspace*{-1mm}

\begin{thebibliography}{33}
\providecommand{\natexlab}[1]{#1}
\providecommand{\url}[1]{\texttt{#1}}
\providecommand{\urlprefix}{URL }
\expandafter\ifx\csname urlstyle\endcsname\relax
  \providecommand{\doi}[1]{doi:\discretionary{}{}{}#1}\else
  \providecommand{\doi}{doi:\discretionary{}{}{}\begingroup
  \urlstyle{rm}\Url}\fi

\bibitem[{Alonso et~al.(2019)Alonso, Pugh, and Dominguez}]{AlonsoEtAl2019}
Alonso, M., Pugh, T., and Dominguez, C. (2019).
\newblock Redesign of a control engineering course using the flipped classroom
and jitt.
\newblock \emph{IEEE Transactions on Education}, 62(3), 179--187.

\bibitem[{Bogunovic and Theis(2022)}]{BogunovicTheis2022}
Bogunovic, L. and Theis, F. (2022).
\newblock Enhancing problem-based learning with just-in-time content delivery
in data-driven engineering disciplines.
\newblock \emph{Educational Technology Research and Development}, 70,
2001--2020.

\bibitem[{Bradbury(2016)}]{bradbury2016attention}
Bradbury, N.A. (2016).
\newblock Attention span during lectures: 8 seconds, 10 minutes, or more?
\newblock \emph{Adv Physiol Educ}, 40, 509--513.

\bibitem[{Bunce et~al.(2010)Bunce, Flens, and Neiles}]{bunce2010long}
Bunce, D.M., Flens, E.A., and Neiles, K.Y. (2010).
\newblock How long can students pay attention in class? a study of student
attention decline using clickers.
\newblock \emph{Journal of Chemical Education}, 87(12), 1438--1443.

\bibitem[{Chang et~al.(2022)Chang, Chen, and Wu}]{ChangEtAl2022}
Chang, Y., Chen, H.T., and Wu, H. (2022).
\newblock Personalized adaptive learning for control systems: A big data and
machine learning approach.
\newblock In \emph{Proceedings of the IEEE International Conference on Advanced
Learning Technologies (ICALT)}, 110--115.

\bibitem[{Cheng and Xu(2020)}]{ChengXu2020}
Cheng, L. and Xu, G. (2020).
\newblock Adaptive e-learning with real-time feedback: A just-in-time teaching
approach.
\newblock \emph{IEEE Access}, 8, 215394--215405.

\bibitem[{Claros and Duart(2021)}]{ClarosDuart2021}
Claros, I. and Duart, J.M. (2021).
\newblock Using formative feedback to improve engineering students’ learning
experiences: An iterative design approach.
\newblock \emph{International Journal of Educational Technology in Higher
Education}, 18, 60.

\bibitem[{Fox and Doherty(2021)}]{FoxDoherty2021}
Fox, J. and Doherty, I. (2021).
\newblock Implementing just-in-time teaching to enhance student engagement in
stem education.
\newblock \emph{Computers \& Education}, 168, 104197.

\bibitem[{Fukuda et~al.(2024)Fukuda, Nesbit, and Winne}]{2024_EffectsJustintimeInquiry_fukuda}
Fukuda, M., Nesbit, J.C., and Winne, P.H. (2024).
\newblock Effects of just-in-time inquiry prompts and principle-based
self-explanation guidance on learning and use of domain texts in
simulation-based inquiry learning.
\newblock \emph{Front. Educ.}, 9, 1446941.

\bibitem[{German Higher~Education(2024)}]{hochschulkompass2023}
German Higher~Education, R.C. (2024).
\newblock Types of higher education institutions.
\newblock
\url{https://www.hochschulkompass.de/en/higher-education-institutions/higher-education-landscape/types-of-higher-education-institutions.html}.
\newblock [Accessed Decembre 31, 2021].

\bibitem[{{Gürdür Broo} et~al.(2022){Gürdür Broo}, Kaynak, and Sait}]{2022_rethink_education}
{Gürdür Broo}, D., Kaynak, O., and Sait, S.M. (2022).
\newblock Rethinking engineering education at the age of industry 5.0.
\newblock \emph{Journal of Industrial Information Integration}, 25, 100311.

\bibitem[{Gupta and Lee(2021)}]{GuptaLee2021}
Gupta, R. and Lee, W. (2021).
\newblock An iot-enabled just-in-time teaching framework for enhancing student
engagement in engineering labs.
\newblock \emph{IEEE Transactions on Learning Technologies}, 14(4), 482--493.

\bibitem[{HasanzadeZonuzy et~al.(2018)HasanzadeZonuzy, Arefizadeh, Talebpour,
Shakkottai, and Darbha}]{2018_Platooning_HasanzadeZonuzy}
HasanzadeZonuzy, A., Arefizadeh, S., Talebpour, A., Shakkottai, S., and Darbha,
S. (2018).
\newblock Collaborative platooning of automated vehicles using variable
time-gaps.
\newblock In \emph{2018 Annual American Control Conference (ACC)}, 6715--6722.

\bibitem[{Jonsson(2015)}]{2015_UsingFlippedClassroom_jonssonc}
Jonsson, H. (2015).
\newblock Using flipped classroom, peer discussion, and just-in-time teaching
to increase learning in a programming course.
\newblock In \emph{2015 {{IEEE Frontiers}} in {{Education Conference}}
({{FIE}})}, 1--9. IEEE, Camino Real El Paso, El Paso, TX, USA.

\bibitem[{Kille et~al.(2024)Kille, Leibold, Karg, Varga, and Hohmann}]{2024_sean_roman}
Kille, S., Leibold, P., Karg, P., Varga, B., and Hohmann, S. (2024).
\newblock Human-variability-respecting optimal control for physical
human-machine interaction.
\newblock In \emph{2024 33rd IEEE International Conference on Robot and Human
Interactive Communication (ROMAN)}, 1595--1602.

\bibitem[{Lindenmaier et~al.(2022)Lindenmaier, Aradi, Bécsi, and Törő}]{2022_sesorFusion_Linden}
Lindenmaier, L., Aradi, S., Bécsi, T., and Törő, O. (2022).
\newblock Comparison of sensor data fusion algorithms for automotive perception
system.
\newblock In \emph{2022 IEEE 20th Jubilee World Symposium on Applied Machine
Intelligence and Informatics (SAMI)}, 000089--000096.

\bibitem[{Makransky and Petersen(2019)}]{MakranskyPetersen2019}
Makransky, G. and Petersen, G.B. (2019).
\newblock Investigating the feasibility of simulation-based labs for control
engineering courses: A focus on student motivation and learning strategies.
\newblock \emph{Virtual Reality}, 23(2), 201--214.

\bibitem[{Oprean et~al.(2011)Oprean, Fabian, Brumar, and Bărbat}]{2011_bounded_rationality}
Oprean, C., Fabian, R.D., Brumar, C.I., and Bărbat, B.E. (2011).
\newblock Bounded rationality for “just in time” education.
\newblock \emph{Procedia - Social and Behavioral Sciences}, 30, 983--987.
\newblock 2nd World Conference on Psychology, Counselling and Guidance - 2011.

\bibitem[{Orr et~al.(2022)Orr, Williams, and Penman}]{OrrEtAl2022}
Orr, R., Williams, M., and Penman, M. (2022).
\newblock Integrating real-world problem-solving in project-based control
engineering courses.
\newblock \emph{Education for Chemical Engineers}, 38, 14--22.

\bibitem[{{Perez-Poch} and Lopez(2017)}]{2017_JustintimeTeachingImproves_perez-poch}
{Perez-Poch}, A. and Lopez, D. (2017).
\newblock Just-in-time teaching improves engagement and academic results among
students at risk of failure in computer science fundamentals.
\newblock In \emph{2017 {{IEEE Frontiers}} in {{Education Conference}}
({{FIE}})}, 1--7. IEEE, Indianapolis, IN.

\bibitem[{Peter and De~Vries(2018)}]{PeterDeVries2018}
Peter, S. and De~Vries, P. (2018).
\newblock Integrating project-based learning and flipped classroom strategies
in control engineering education.
\newblock \emph{International Journal of STEM Education}, 5, 23.

\bibitem[{Rossiter et~al.(2023)Rossiter, Cassandras, Hespanha, Dormido, de~la
Torre, Ranade, ..., and Parisini}]{Rossiter2023}
Rossiter, J.A., Cassandras, C.G., Hespanha, J.P., Dormido, S., de~la Torre, L.,
Ranade, G., ..., and Parisini, T. (2023).
\newblock Control education for societal-scale challenges: A community roadmap.
\newblock \emph{Annual Reviews in Control}, 55.

\bibitem[{Rossiter et~al.(2018)Rossiter, {Pasik-Duncan}, Dormido, Vlacic,
Jones, and Murray}]{2018_SurveyGoodPractice_rossiter}
Rossiter, J.A., {Pasik-Duncan}, B., Dormido, S., Vlacic, L., Jones, B., and
Murray, R. (2018).
\newblock A survey of good practice in control education.
\newblock \emph{European Journal of Engineering Education}, 43(6), 801--823.

\bibitem[{Rossiter et~al.(2019)Rossiter, {Pasik-Duncan}, Visioli, Serbezov,
Zakova, Huba, and Dormido}]{2019_TechnicalCommitteeControl_rossiter}
Rossiter, J., {Pasik-Duncan}, B., Visioli, A., Serbezov, A., Zakova, K., Huba,
M., and Dormido, S. (2019).
\newblock Technical {{Committee}} on {{Control Education}} [{{Technical
Activities}}].
\newblock \emph{IEEE Control Syst.}, 39(4), 20--22.

\bibitem[{Schwerdt and Wuppermann(2011)}]{schwerdt2011traditional}
Schwerdt, G. and Wuppermann, A.C. (2011).
\newblock Is traditional teaching really all that bad? a within-student
between-subject approach.
\newblock \emph{Economics of Education Review}, 30(2), 365--379.

\bibitem[{Smith et~al.(2019)Smith, Johnson, and Lee}]{SmithEtAl2019}
Smith, T., Johnson, M., and Lee, K. (2019).
\newblock Personalized learning in large-scale online courses: A framework for
implementation and evaluation.
\newblock \emph{Educational Technology Research and Development}, 67(1),
57--73.

\bibitem[{Tucker and Griffin(2017)}]{TuckerGriffin2017}
Tucker, B. and Griffin, P. (2017).
\newblock A just-in-time teaching model for hybrid learning environments:
Bridging the gap between theory and practice.
\newblock \emph{Innovations in Education and Teaching International}, 54(4),
376--388.

\bibitem[{Umezawa et~al.(2018)Umezawa, Hashimoto, and Shiozawa}]{UmezawaEtAl2018}
Umezawa, T., Hashimoto, A., and Shiozawa, N. (2018).
\newblock Analyzing the effects of immediate feedback on students’ engagement
in remote labs.
\newblock In \emph{Proceedings of the IEEE Global Engineering Education
Conference (EDUCON)}, 1083--1087.

\bibitem[{Varga(2024)}]{varga2024toward}
Varga, B. (2024).
\newblock Toward adaptive cooperation: Model-based shared control using
lq-differential games.
\newblock \emph{Acta Polytechnica Hungarica}, 21(10).

\bibitem[{Varga et~al.(2021)Varga, Inga, Lemmer, and Hohmann}]{2021_OrdinalPotentialDifferential_varga}
Varga, B., Inga, J., Lemmer, M., and Hohmann, S. (2021).
\newblock Ordinal {{Potential Differential Games}} to {{Model Human-Machine
Interaction}} in {{Vehicle-Manipulators}}.
\newblock In \emph{2021 {{IEEE Conference}} on {{Control Technology}} and
{{Applications}} ({{CCTA}})}, 728--734. {IEEE}, {San Diego, CA, USA}.

\bibitem[{Varga et~al.(2023{\natexlab{a}})Varga, Yang, and Hohmann}]{varga2023intention}
Varga, B., Yang, D., and Hohmann, S. (2023{\natexlab{a}}).
\newblock Intention-aware decision-making for mixed intersection scenarios.
\newblock In \emph{2023 IEEE 17th International Symposium on Applied
Computational Intelligence and Informatics (SACI)}, 000369--000374. IEEE.

\bibitem[{Varga et~al.(2023{\natexlab{b}})Varga, Yang, Martin, and Hohmann}]{varga2023cooperative}
Varga, B., Yang, D., Martin, M., and Hohmann, S. (2023{\natexlab{b}}).
\newblock Cooperative decision-making in shared spaces: Making urban traffic
safer through human-machine cooperation.
\newblock In \emph{2023 IEEE 21st Jubilee International Symposium on
Intelligent Systems and Informatics (SISY)}, 000109--000114. IEEE.

\bibitem[{Visioli(2023)}]{2023_TC_IFAC}
Visioli, A. (2023).
\newblock Control education: Tc 9.4 developments and vision.
\newblock \emph{IFAC-PapersOnLine}, 56(2), 332--335.
\newblock 22nd IFAC World Congress.


\end{thebibliography}

\end{document}